\documentclass[aps,twocolumn,a4paper,showpacs,amsmath,amssymb]{revtex4}

\usepackage{amssymb}
\usepackage{enumerate}
\usepackage{graphicx}
\usepackage{hyperref}
\usepackage[a4paper,left=2cm,right=2cm,top=2cm,bottom=2cm]{geometry}

\begin{document}

\title{Characterization of Subgraphs Relationships and \\ Distribution in Complex Networks}

\author{Lucas Antiqueira}
\email{lantiq@gmail.com}
\author{Luciano da Fontoura Costa}
\email{luciano@if.sc.usp.br}
\affiliation{Instituto de F\'{\i}sica de S\~{a}o Carlos, Universidade de S\~{a}o Paulo, Av. Trabalhador S\~{a}o Carlense~400, Caixa Postal~369, CEP 13560-970, S\~{a}o Carlos, S\~ao Paulo, Brazil}


\begin{abstract}
A network can be analyzed at different topological scales, ranging from single nodes to motifs, communities, up to the complete structure. We propose a novel intermediate-level topological analysis that considers non-overlapping subgraphs (connected components) and their interrelationships and distribution through the network.  Though such subgraphs can be completely general, our methodology focuses the cases in which the nodes of these subgraphs share some special feature, such as being critical for the proper operation of the network. Our methodology of subgraph characterization involves two main aspects: (i)~a distance histogram containing the distances calculated between all subgraphs, and (ii)~a merging algorithm, developed to progressively merge the subgraphs until the whole network is covered. The latter procedure complements the distance histogram by taking into account the nodes lying between subgraphs, as well as the relevance of these nodes to the overall interconnectivity. Experiments were carried out using four types of network models and four instances of real-world networks, in order to illustrate how subgraph characterization can help complementing complex network-based studies.
\end{abstract}


\maketitle


\section{Introduction}

Because of their flexibility to represent, model and simulate
virtually any discrete structure, complex networks 
\cite{Albert2002,Dorogovtsev2002,Newman2003,Boccaletti2006,Costa2005a}
have been extensively studied and applied to the most diverse 
problems~\cite{Costa2007c}, ranging from transportation (e.g. 
flights~\cite{Barabasi2007}) to communications (e.g. 
Internet~\cite{Faloutsos1999}). Complex networks are `complex' because they 
exhibit particularly intricate and heterogeneous connectivity, e.g. by 
involving hubs~\cite{Faloutsos1999,Barabasi1999} or 
communities~\cite{Clauset2004,Newman2006}.  As shown 
recently~\cite{Costa2004b,Costa2008}, most real-world complex networks also 
include in their structure regular patches of connectivity, i.e. subgraphs 
whose nodes present similar topological measurements.  All in all, the 
heterogeneity of complex networks tends to range along several topological 
scales, extending from the individual node level through mesoscopic structures 
such as modules and regular subgraphs, up to the whole network level.  As a 
matter of fact, it is precisely the heterogeneous distribution of structural
features along the several scales which defines the intricate
organization and most interesting structural and dynamical properties
of complex networks.

Although several works have investigated mesoscopic features of the
connectivity of complex networks, e.g. by considering their respective
communities \cite{Clauset2004,Newman2006} and/or paths between
different portions of the networks \cite{Lopez2007,Costa2008a}, few
works (e.g.~\cite{Makarov2005,Costa2008}) have focused on the study
and characterization of the \emph{distribution} of nodes and subgraphs
within a given network. Such nodes and subgraphs of special interest
arise in several situations, not only as communities or regular
patches, but also with respect to extreme values of specific
topological measurements.  For instance, the nodes (or edges) with
betweenness centrality higher than a given threshold can give rise to
several subgraphs inside a network.  It should be observed that such
subgraphs, investigated in this work, do not necessarily yield a
partition of the original network, as typically they do not encompass
all the original nodes.  At the same time, these subgraphs are
henceforth assumed to be connected components and not to overlap one
another.

Given a set of disjoint subgraphs of a network expressing specific
properties of interest, it becomes critically important to
characterize how these subgraphs are distributed through the network,
as such an information can be particularly important regarding the
overall organization of the network and its dynamics. Going back to
the above example with betweenness centrality, if highly central 
subgraphs are found to be close one another (in terms of shortest path
length between them), the portion of the original network containing
such subgraphs can be understood as corresponding to a critical
bottleneck for the whole system under analysis.  On the other hand, a
more uniform distribution of subgraphs with highest betweenness
centrality suggests a system less critically structured for
communications.  Several similar situations can be characterized with
respect to other types of subgraphs, including other measurements as
well as communities and regular patches.  Yet, few works have
addressed the specific issue of how critical nodes or subgraphs are
distributed through the network topology.

\begin{figure}[!htb]
  \centering
  \includegraphics[width=1.0\columnwidth]{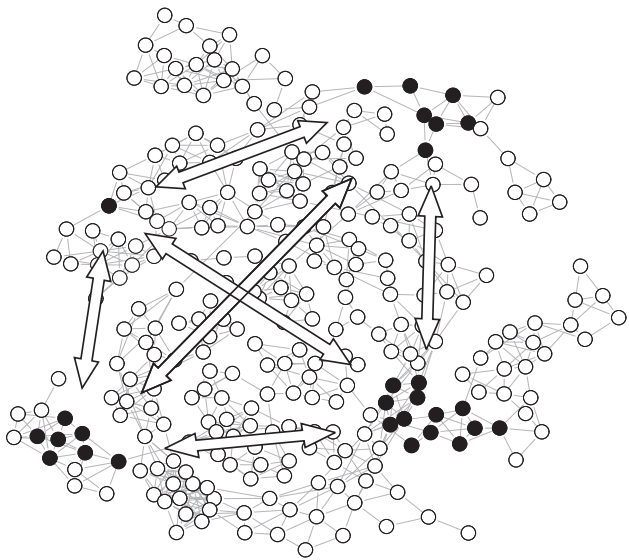}
    \caption{A graph containing four subgraphs whose nodes are highlighted. The
approach reported in this article is aimed at characterizing the topological
distribution and relationships between these subgraphs.}
    \label{fig:intro}
\end{figure}

The objective of the present work is to develop and apply a
comprehensive framework for characterization of the distribution of
subgraphs of specific interest within a given complex network.  In
order to do so, we resource to the distances, quantified in terms of
the shortest path lengths, between each pair of given subgraphs.  Such
distances are organized into a histogram, which can provide valuable
information about the topological distribution of the subgraphs.  For
instance, a sharp peak in such a histogram at a small value of
distance will indicate that the subgraphs are all close one another.
Though the distances between subgraphs provide valuable information
about their overall distribution, it is also interesting to have the
means for progressively merging subgraphs in order to obtain connected
components incorporating the critical regions.  Therefore, we also
report an algorithm which allows the progressive merging of the
subgraphs in terms of successive distance values, up to the point of
containing all the given subgraphs. This merging is based on the
morphological operation called \emph{dilation}. Figure~\ref{fig:intro} 
depicts a network with four subgraphs and some bidirectional arrows which 
denote the distance and merging relationships we want to characterize.
The potential of the distance histograms and subgraph merging 
algorithm are illustrated with respect to both theoretical as 
well as real-world complex networks, including the Barab\'{a}si-Albert 
model \cite{Barabasi1999} as well as the power grid of the western 
states of the USA \cite{Watts1998}.

This article starts by presenting the basic concepts and methods
(Section~\ref{sec:basic}) and proceeds by describing the distance
histogram and merging algorithm (Section~\ref{sec:charac}), which are
then illustrated with respect to theoretical and real-world networks
(Section~\ref{sec:results}).


\section{Basic Concepts and Methods}
\label{sec:basic}

A network can be represented by a graph $G(V,E)$, where
$V=\{v_1,v_2,\ldots,v_N\}$ is its set of $N$ vertices (or nodes) and
$E=\{e_1,e_2,\dots,e_L\}$ is its set of $L$ edges (or links). An edge
$e_k$ is a pair $(v_i,v_j)$ that represents a connection between nodes
$v_i \in V$ and $v_j \in V$. The set of edges $E$ can be encoded into
an adjacency matrix $A$, of dimension $N\times{}N$, with elements
$A(i,j) = 1$ whenever there is an edge from node $i$ to node $j$, with
$A(i,j) = 0$ being imposed otherwise. Notice that $A$ is symmetric for
undirected graphs.

In what follows we present the definitions of the adopted measurements
for undirected graphs, as well as the details of the considered
artificial and real-world networks.

\subsection{Network Measurements}
\label{sec:measur}

The network measurements reviewed in this section have been frequently
employed in the field of complex networks. For more details, please
refer to the review article \cite{Costa2005a}.

\setcounter{paragraph}{0}

\paragraph{Degree:}
The node degree $k(i)$ corresponds to the number of edges attached to
a node~$i$. Using the adjacency matrix $A$, the degree can be obtained
by:
\begin{equation}
k(i) = \sum_{j=1}^{N}{A(i,j)} .
\end{equation}

\paragraph{Clustering Coefficient:}
This measurement reflects the density of connections between the
neighbors of a node $i$. Let: \[ \eta(i) = \{j\ |\ A(i,j)=1, i \neq j\}
\] be the set of neighbors of $i$, and: \[
\epsilon(i) =
\sum_{u,v\in{}\eta(i), u\neq{}v}{A(u,v)} \] be the number of edges
between the neighbors of $i$. The clustering coefficient of a node $i$
is defined as:
\begin{equation}
c(i) = \frac{2\epsilon(i)}{\left|\eta(i)\right| \left(
\left|\eta(i)\right|-1 \right)} ,
\end{equation}
where $\left|\eta(i)\right|$ is the cardinality of $\eta(i)$, i.e. it
is the number of neighbors of $i$.

\paragraph{Length of Shortest Paths:}
The proximity between nodes is usually quantified in terms of shortest
paths. A path $p(i,j)$ extending from node $i$ to node $j$ is denoted
by a sequence of neighboring nodes: \[ p(i,j) = (v_1, v_2, \ldots, v_l,
v_{l+1}) \] where $A(v_i,v_{i+1})=1$, $v_1=i$, $v_{l+1}=j$ and the
length of the path is $\omega(p(i,j)) = l$. Notice that the length of
a path is the number of edges along it. The length of the shortest
path between two nodes $i$ and $j$ is thus given by:
\begin{equation}
	s(i,j) = \min\{\omega(p(i,j))\},
	\label{eq:spath}
\end{equation}
which is the minimum amount of steps (edges) needed to reach node $j$
after starting at node $i$ (or vice versa in the case of undirected
networks).

\paragraph{Betweenness Centrality:}
This measurement is closely related to the shortest path. Consider the
set: \[ \sigma(i,j) = \{p(i,j)\ |\
\omega(p(i,j))=s(i,j)\} \] of all shortest paths between $i$ and
$j$. Moreover, the set: \[ \sigma(i,v,j) = \{p(i,j)\ |\
p(i,j)\in{}\sigma(i,j) \textrm{~and~} v\in{}p(i,j) \} \] contains all
the shortest paths between $i$ and $j$ that pass through node $v$. The
betweenness centrality of a node $v$ is given as:
\begin{equation}
b(v) = \sum_{i\neq{}j}{\frac{\left| \sigma(i,v,j) \right|}{ \left| \sigma(i,j) \right|}},
\end{equation}
which takes the sum over all possible pairs of distinct nodes $i$ and
$j$. Informally speaking, this centrality measurement quantifies the
participation of $v$ in minimum paths.

\subsection{Network Models}
\label{sec:models}
\setcounter{paragraph}{0}

Four theoretical network models were chosen in this work in order to
construct undirected networks. For each model, 100 realizations
(networks) were performed with $N=1,000$ nodes and mean degree
$\left\langle k
\right\rangle = 6$. The general characteristics of these models are
given below:

\paragraph{Erd\H{o}s-R\'{e}nyi (ER):}
In this model, every possible pair of nodes $(i,j)$ is connected with
uniform probability $p$ \cite{Erdos1959}. For an ER network, the mean node
degree is given by $\left\langle{}k\right\rangle = p(N-1)$ in the
large network limit $N \rightarrow \infty$. Moreover, this model
yields random networks with a Poisson degree distribution, which
implies a characteristic mean degree, i.e. the node degrees do not
greatly deviate from $\left\langle{}k\right\rangle$.

\paragraph{Watts-Strogatz (WS):}
The WS model generates networks exhibiting the small-world property,
i.e. high average clustering coefficient and low average shortest
paths \cite{Watts1998}. In order to obtain a WS network, we start with
a regular ring-shaped network with $N$ nodes, where every node is
connected to its $\kappa$ nearest neighbors in both directions. Then,
each edge is moved (rewired) to another position with probability
$p$. Depending on $p$, an ER network can approach the features of the
initial regular network (for $p \rightarrow 0$) or of a random network
(for $p \rightarrow 1$). In our experiments, we employed $p = 0.2$.
Also notice that the mean node degree of a WS network is 
$\left\langle{}k\right\rangle = 2\kappa$.

\paragraph{Barab\'{a}si-Albert (BA):}
Networks with a power-law degree distribution can be obtained by
considering the BA model \cite{Barabasi1999}. This type of network
contains a few nodes, called hubs, concentrating many connections,
while the majority of nodes have only a few links. A BA network is
generated by adding new nodes to an initial network of $m_0$
nodes. Each newly added node is connected to $m$ previous nodes, with
the probability of connections being proportional to the respective
degrees. The average node degree of a BA network is
$\left\langle{}k\right\rangle = 2m$.

\paragraph{Geographical (GG):}
In contrast to the ER, WS and BA models, a geographical model
considers the spatial position of nodes to create edges
\cite{Costa2005a}, so that the spatial adjacency between nodes often
strongly influences the respective connectivity. In the geographical
model adopted here (called GG), randomly placed nodes are distributed
through a bi-dimensional grid of size $L \times L$, and edges are
established among nodes geographically close to each other,
i.e. separated by a distance not greater than $R$. Thus, long-range
connections are not created, implying longer paths than in the
previous models presented in this section. The mean degree of a GG
network can be estimated as $\left\langle{}k\right\rangle \approx
{\pi{}R^2N}/{L^2}$.

\subsection{Real-World Networks}
\label{sec:realnets}
\setcounter{paragraph}{0}

A set of real networks, as described below, has also been used in our
experiments. These networks have been chosen so as to provide a
representative sample of several types of real-world networks of
general interest.  Table~\ref{tab:realnets} provides a quick reference
with basic information about each network.

\paragraph{Co-authorship in Network Science:} 
This network, called NetScience, expresses the co-authorship
relationships between scientists that published papers in the field of
complex networks \cite{Newman2006a}. It was compiled by M.E.J. Newman
in May 2006 \footnote{The co-authorship network is available at the
website of M.E.J.~Newman: \\
http://www-personal.umich.edu/\~{}mejn/netdata.}  using the
references cited in two surveys of the field
\cite{Newman2003,Boccaletti2006}, plus some manually added
references. Each scientist is a node in this network, while an
undirected edge is created between two scientists whenever they have
published at least one joint paper. NetScience has $1,589$ nodes, of
which 379 are inside the largest connected component, which is the
part we used in our experiments. Henceforth, whenever we mention
NetScience, we refer to its largest connected component. Moreover, we
do not take into consideration the original weights of this network.

\paragraph{Email Communications:}
We also considered a graph reflecting the flow of email messages
exchanged among the members of the University at Rovira i Virgili
(Spain) \cite{Guimera2003}. This network, compiled in the research
group of A.~Arenas \footnote{The email network is available at the
website of A.~Arenas: \\
http://deim.urv.cat/\~{}aarenas/data/welcome.htm.}, 
has a single connected component, where each email address is
identified by a node (there are $N=1,133$ addresses), and a message
sent from node $i$ to node $j$ is represented by an undirected
unweighted edge $(i,j)$. The authors removed bulk emails,
i.e. messages sent to more than 50 addresses, before defining the
edges in this network.

\paragraph{Power Grid:}
This network represents the topology of the power grid of the western
states of the USA \cite{Watts1998}, and was compiled by D.~Watts and
S.~Strogatz \footnote{The power grid network is available at the
website of D.~Watts: \\
http://cdg.columbia.edu/cdg/datasets.}. A power grid is the
structure that underlies the transmission of electricity from power
plants to consumers. The power grid of the USA western states is a
single connected component with $4,941$ nodes interconnected by
undirected and unweighted links.

\paragraph{Internet-AS:} The connections that associate ASs 
(Autonomous Systems) in the Internet were considered by M.E.J.~Newman
in the compilation of this network \footnote{The Internet snapshot is
available at the website of M.E.J.~Newman: \\
http://www-personal.umich.edu/\~{}mejn/netdata.}. An AS is a
group of computer networks that share the same routing policy and have a
centralized administration. Using BGP (Border Gateway Protocol) data
of July 22, 2006, Newman reconstructed the links between $22,963$ ASs,
which yielded a connected graph with unweighted and undirected
edges. Due to the nature of BGP, which is a gateway protocol used to
route data packets between ASs, it was possible to retrieve information
about the physical links of the Internet at the AS level.

\begin{table}[!htb]
	\caption{Basic information about the real networks used in our experiments. For each network, we show its number of nodes $N$, number of edges $L$ and respective mean degree $\left\langle k \right\rangle$.}
	\centering
	\small
	\begin{tabular}{l@{\hspace{1.5em}}r@{\hspace{1.5em}}r@{\hspace{1.5em}}r}
	\hline \hline
	 & $N$ & $L$ & $\left\langle k \right\rangle$ \\
	\hline
	NetScience  & $379$    & $914$    & $4.82$ \\ 
	Email       & $1,133$  & $5,451$  & $9.62$ \\ 
	Power Grid  & $4,941$  & $6,594$  & $2.67$ \\ 
	Internet-AS & $22,963$ & $48,436$ & $4.22$ \\ 
	\hline \hline
	\end{tabular}
	\label{tab:realnets}
\end{table}


\section{Characterization of Subgraphs}
\label{sec:charac}

\begin{figure}[!htb]
  \centering
  \includegraphics[width=0.9\columnwidth]{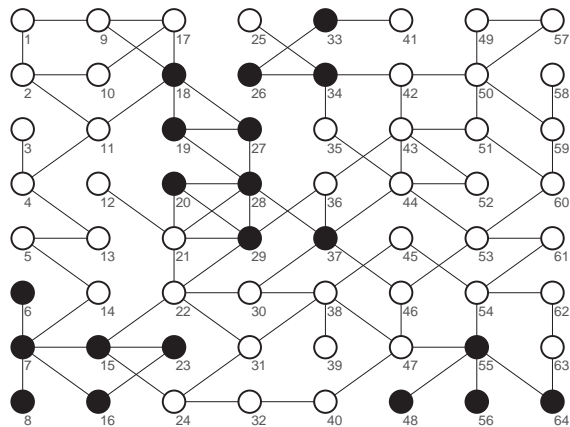}
    \caption{Graph $G=(V,E)$ with $N=64$ nodes, where each node is
    identified by a number $v$. This graph has four subgraphs $G_{1}$,
    $G_{2}$, $G_{3}$ and $G_{4}$, whose set of nodes are
    $V_{1}=\{6,7,8,15,16,23\}$, $V_{2}=\{18,19,20,27,28,29,37\}$,
    $V_{3}=\{26,33,34\}$ and $V_{4}=\{48,55,56,64\}$. These subgraphs
    correspond to the four connected components with black nodes.}
    \label{fig:net}
\end{figure}

The method for analyzing subgraph interconnectivity introduced in this
section assumes that the graph/network $G(V,E)$ under study is
undirected, unweighted and connected. It also requires that $C$
subgraphs $G_i=(V_i,E_i)$, $1 \leq i \leq C$, be defined such that:
\begin{enumerate}[(i)]
	\setlength{\parskip}{0pt}
	\item $V_i \subset V$,
	\item $V_i\neq\varnothing$,
	\item $V_i\cap V_j = \varnothing$ for every $i \neq j$,
	\item $(v_i,v_j) \in E_i$ if and only if $v_i \in V_i$, $v_j \in V_i$ and $(v_i,v_j) \in E$,
	\item $G_i$ is a connected component,
	\item Two different subgraphs $G_i$,$G_j$ are not direct neighbors.
\end{enumerate}
In order to create a subgraph $G_i$, it is enough to define a valid
$V_i$, since $E_i$ must contain all the edges of $E$ that connect
pairs of nodes included in $V_i$ (rule~(iv)). We refer to a `valid'
$V_i$ as a non-empty subset of $V$ (rules~(i) and~(ii)) that results
in a connected component (rule~(v)) not sharing any nodes or edges
with other subgraphs (rules~(iii) and~(vi)). Furthermore, the
selection of nodes for subgraphs $G_i$ also depends on the specific
study being carried out, given that the above conditions are
followed. Figure~\ref{fig:net} shows an example of a graph $G$ with $N
= 64$ and four subgraphs $G_1,\ldots,G_4$.

\subsection{Distance Histogram}
\label{sec:hist}

One way to analyze subgraph interrelationship is by computing the
distance between every pair of subgraphs. We define this distance to
correspond to the length of the shortest path between subgraphs and,
therefore, (\ref{eq:spath}) can be used for this purpose. Notice that
there are at least $|V_i||V_j|$ different paths between two different
subgraphs $G_i$ and $G_j$, because each path may start at any node of
the source subgraph and end at any node of the destination
subgraph. The length of the shortest path between two different
subgraphs can now be defined as:
\begin{equation}
s(G_i,G_j) = \mathop{\min}_{w_i \in V_i, w_j \in V_j}\left\{ s(w_i,w_j) \right\}-1,
\end{equation}
imposing that $s(G_i,G_j) = 0$ whenever $i=j$. Notice that the length
of the shortest path is decremented by one, which changes the length
of the distance from an edge-orientation to a node-orientation. This
modification has been done because there must be at least one node, or
two edges, between a pair of subgraphs (i.e. the distance would start
at two). Thus, a node-oriented distance is preferred because it is
more intuitive. We therefore define the matrix $D_s$ of order $C
\times C$ with elements $D_s(i,j) = s(G_i,G_j)$, i.e. it encodes every
distance between all $C$ subgraphs of $G$. As an example, the matrix
$D_s$ for the four subgraphs in Figure~\ref{fig:net} is given as:
\[
D_{s}=\left[
\begin{array}{cccc}
0 & 1 & 5 & 4\\
1 & 0 & 2 & 2\\
5 & 2 & 0 & 5\\
4 & 2 & 5 & 0
\end{array}
\right]
\]
where $D_s$ is symmetric because $G$ is an undirected graph. If these
distances (excluding the diagonal of $D_s$) are placed in a histogram,
the overall proximity between subgraphs can be examined more easily
than just observing $D_s$, as will become clearer in the experiments
reported in Section~\ref{sec:results}.

\subsection{Subgraph Merging}
\label{sec:expansion}

The method detailed in this section aims at gradually merging
subgraphs $G_{i}$ inside the original graph $G$, while giving special
attention to the relationship between them. We implement this
progressive merging, or expansion, in terms of the gradual growth of
the subgraphs $G_i$ towards graph $G$, which is accomplished by adding
to the subgraphs $G_i$ nodes of $G$ that do not belong to any $G_i$
yet (and also adding the necessary edges, as specified in the
definition of the subgraphs $G_i$ presented earlier in this paper).

In order to achieve a gradual subgraph merging, some vertices of $G$
need to be included in the expansion earlier than others. In our
methodology, higher relevance is given to the nodes inside a short
path connecting some pair of different subgraphs $(G_{i}$,$G_{j})$. In
this manner, the merging is controlled by the length of paths between
subgraphs. More specifically, for a node~$v$ outside every subgraph
$G_i$, we compute the length of the minimum path between all pairs of
different subgraphs $(G_{i}$,$G_{j})$ that necessarily pass through
node~$v$. This is understood as the \emph{relevance} of a node $v$ in
the merging of subgraphs. In other words, a node that is close to only
one subgraph is considered a member of a weak tie, because it does not
take part in short paths connecting that subgraph with others.

\begin{figure}[!htb]
  \centering
  \includegraphics[width=0.9\columnwidth]{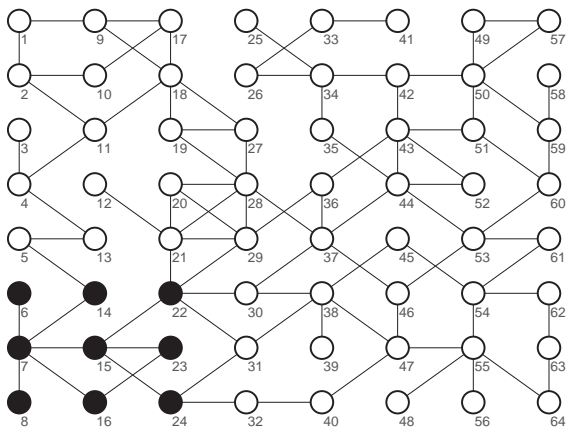}
  \caption{Dilation $\delta(G_1)$ of the subgraph $G_1$ of Figure~\ref{fig:net}. $G_1$ has nodes $V_{1}=\{6,7,8,15,16,23\}$, and the dilation is formed by nodes $V_{1} \cup \{14,22,24\}$, represented in black in the figure.}
  \label{fig:net_dilation}
\end{figure}

The aforementioned merging can be carried out by applying consecutive
dilations \cite{Heijmans1992,Costa2004b} in the subgraphs of $G$. The
dilation is a morphological operation $\delta(g)$, defined over a
subgraph $g$ of $G$, that yields another subgraph that is equal to the
union of the original subgraph and its neighbors in $G$ (plus the
respective edges). Figure~\ref{fig:net_dilation} illustrates the
dilation of the subgraph $G_1$ of Figure~\ref{fig:net}, which is
formed by vertices $V_{1}=\{6,7,8,15,16,23\}$. The dilation
$\delta(G_1)$ results in a subgraph with nodes $V_{1} \cup
\{14,22,24\}$ (i.e. nodes
\{14,22,24\} are neighbors of $G_1$), along with the respective edges
that connect these nodes in $G$.

Dilations are employed as an intermediate step in our method. When
$\delta(G_{i})$ is applied sequentially and recursively inside $G$,
until no more dilations are possible, a distance map is traced between
the subgraph $G_i$ and the other nodes of $G$. This recursive dilation
is denoted by:
\begin{equation}
\delta_d(G_{i}) = \delta\underbrace{((\ldots(G_{i})\ldots))}_{d \textrm{~times}},
\end{equation}
and the nodes that are included in $\delta_d(G_{i})$, but not in
$\delta_{d-1}(G_{i})$, are said to be at distance $d$, in number of
edges, from $G_{i}$. For $d=0$, the recursive dilation is defined as
$\delta_0(G_{i}) = G_{i}$, and because the dilation
$\delta_{-1}(G_{i})$ is not possible, the nodes inside $G_i$ are
naturally defined to be at distance $0$ from $G_i$. It is worth
pointing out that these distances are different from those given in
the previous section (from matrix $D_s$), which are only calculated
between subgraphs, not between a subgraph and every node of the
network.

Since we are going to dilate all $C$ subgraphs, it is necessary to
apply the dilation $\delta_d(G_{i})$ without considering the nodes of
other subgraphs $G_{j}$, $i\neq{}j$. This particular behavior is
required because the merging is made outwards the set of subgraphs,
and thus it is not necessary to consider nodes that already belong to
a subgraph. Therefore, when dilating a subgraph $i$, some other
subgraph $j$ may block the accessibility of $i$ to some nodes in the
graph. For example, in Figure~\ref{fig:net}, subgraph $G_2$ can not
communicate with nodes 25 and~41 because subgraph $G_3$ is blocking
its access to these nodes. In fact, in this example, only subgraph
$G_3$ is able to communicate with nodes 25 and~41, or, conversely,
these nodes can only access subgraph $G_3$. In what follows, we define
the set of subgraphs accessible from a node $v$ as:
\begin{equation}
Q(v) = \{G_{i}\ |\ v \in \delta_d(G_{i}), \textrm{~for~} 0 \leq d < \infty \},
\end{equation}
and the total number of subgraphs accessible from a node $v$ is:
\begin{equation}
q(v)=\left| Q(v) \right|,
\end{equation}
where $\left| Q(v) \right|$ is the cardinality of $Q(v)$. Notice that
$q(v) \geq 1 $ for any $v$, due to the fact that $G$ is a
connected graph.

The complete set of dilations $\delta_d(G_{i})$, which takes into
account every subgraph $G_{i}$ and every possible dilation starting
from $d=1$ (until there is no more nodes to be added by the dilation),
allows the definition of a distance matrix $D_{\delta}$, of order $N
\times C$. An element $D_{\delta}(v,i)$ of $D_{\delta}$ indicates the
distance between node $v$ and subgraph $G_{i}$. As observed
before, this distance is given by the dilation in number of 
edges. There is an exception requiring special treatment:
for a node $v$ not accessible from a subgraph $G_{i}$, i.e. $G_{i}
\notin Q(v)$, we define $D_{\delta}(v,i) = d_{max}+1$, where
$d_{max}=N-1$ is the maximum possible distance between two nodes in a
graph with $N$ nodes.

The definition of the matrix $D_{\delta}$ is the last step before
specifying a relevance value for each vertex. The shortest path
between any two subgraphs $G_{i}$ and $G_{j}$, $i \neq j$, that
necessarily pass through node $v$, is then defined as the relevance
$r(v)$ of node $v$. Consequently, the lower $r(v)$, the higher is the
relevance of node $v$ in the merging of the subgraphs. More formally,
$r(v)$ is given by:
\begin{widetext}
\begin{equation}
r(v) = 
	\left\{ {
	\begin{array}{lr}
	\mathop{\min}\limits_{1\leq{}i,j\leq{}C,i\neq{}j} \left\{ D_{\delta}(v,i)+D_{\delta}(v,j) \right\}-1, & \mathrm{if}~q(v) > 1 \\
	\mathop{\min}\limits_{1\leq{}i\leq{}C} \left\{ D_{\delta}(v,i) \right\}, & \mathrm{if}~q(v) = 1 \\
	\end{array}
	} \right.	
	\label{eq:relevance}
\end{equation}
\end{widetext}
where the second case is an exception that occurs when node $v$ can
access only one subgraph (or it is inside a subgraph), thus $r(v)$ is
equal to $D_{\delta}(v,i)$, where $i$ is the index of the only
subgraph to which $v$ is connected. In this case, $r(v)$ is solely
controlled by the consecutive dilations of a single subgraph. Observe
also that, in the first case, the minimum expression is decreased by
one. This decreasing scheme was included because otherwise $r(v)$
would always be greater than one for nodes with $q(v) > 1$, an odd
behavior for a quantity that starts at zero. In other words, the
relevance was changed from an edge-oriented to a vertex-oriented one,
similarly to what was done with matrix $D_s$ in the computation of the
distance histogram (Section~\ref{sec:hist}). Therefore,
(\ref{eq:relevance}) allows relevance values greater or equal than one
for every node outside subgraphs $G_i$, while zero relevance is
reserved for the nodes inside some subgraph.

\begin{figure}[!htb]
  \centering
  \includegraphics[width=0.9\columnwidth]{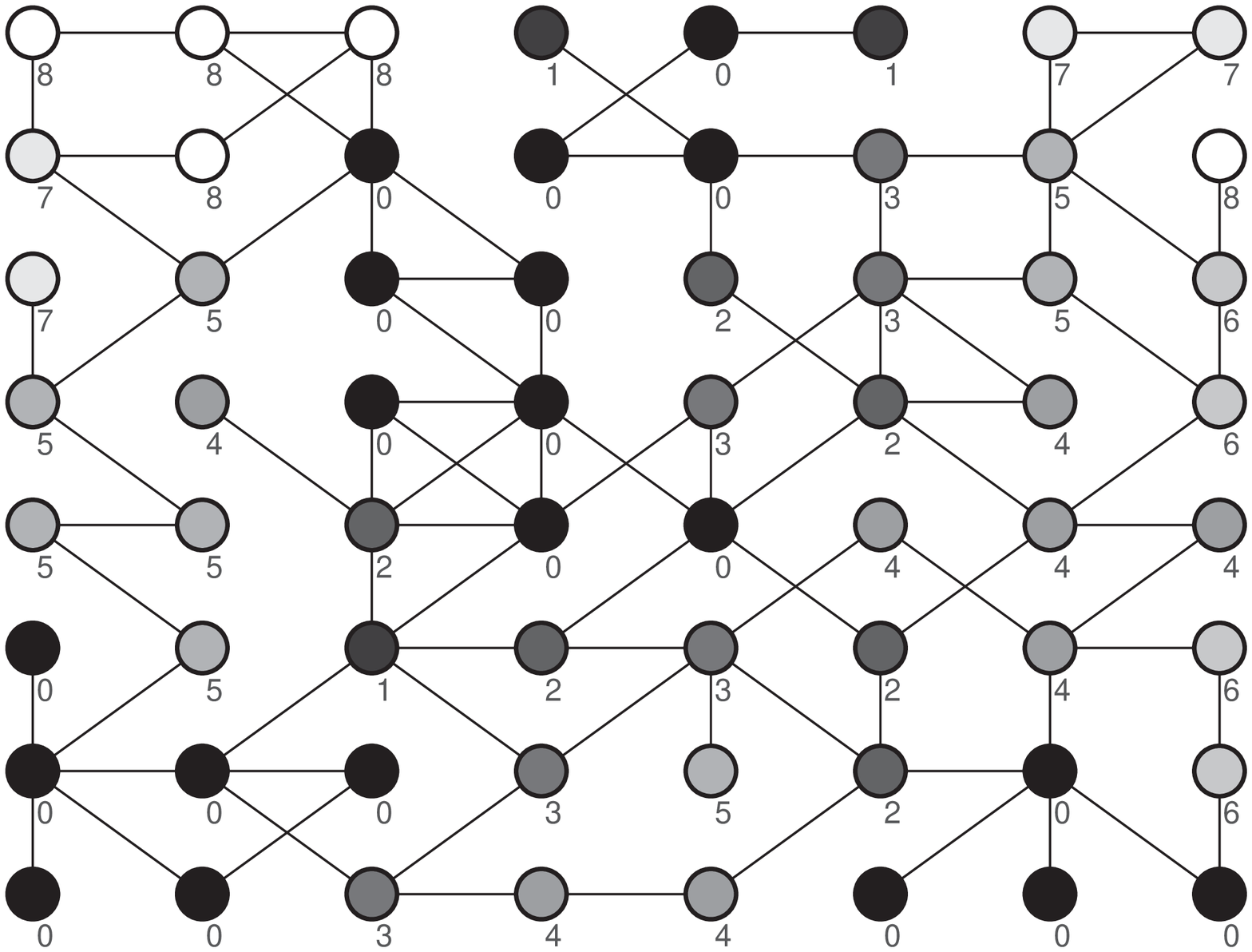}~
  \includegraphics[width=0.08\columnwidth]{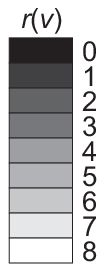}
  \caption{Values of $r(v)$ for each node in the graph $G$ of
  Figure~\ref{fig:net}. The number next to each node denotes $r(v)$
  (please, refer to Figure~\ref{fig:net} for the number $v$ of each
  node). The color of each node is derived from a gray scale ranging
  from black ($r(v)=0$) to white ($r(v)=8$).}  \label{fig:net_relev}
\end{figure}

Figure~\ref{fig:net_relev} shows $r(v)$ for every node of the graph
illustrated in Figure~\ref{fig:net}. In this example, $r(v)$ is shown
both numerically and graphically. The latter approach uses a gray
scale proportional to $r(v)$, ranging from black (when $r(v)=0$, which
refers to the more relevant nodes) to white (when $r(v)=8$,
representing the less relevant nodes in this example). Notice that the
darkest nodes are placed in the shorter paths that connect subgraphs
$G_1,\ldots,G_4$. The nodes of these subgraphs correspond to those
with $r(v)=0$.

Finally, the \emph{merging} of the subgraphs of $G$ is performed by
thresholding the relevance values as follows:
\begin{equation}
V^+ = \left\{ v\ |\ r(v) \leq T \right\},
\end{equation}
where $T\geq0$ is an integer threshold. In addition, $C^+$ subgraphs
$G^+_i = (V^+_i,E^+_i)$, $1 \leq i \leq C^+$, are created such that:
\begin{enumerate}[(i)]
	\setlength{\parskip}{0pt}
	\item $V^+_1 \cup V^+_2 \cup \ldots \cup V^+_{C^+} = V^+$,
	\item $V^+_i\cap V^+_j = \varnothing$ for every $i \neq j$,
	\item $(v_i,v_j) \in E^+_i$ if and only if $v_i \in V^+_i$, $v_j \in V^+_i$ and $(v_i,v_j) \in E$,
	\item $G^+_i$ is a connected component,
	\item Two different subgraphs $G^+_i$,$G^+_j$ are not direct neighbors.
\end{enumerate}
These rules are similar to those given in the definition of the
original subgraphs $G_i$, with the difference that the merged
subgraphs are restricted to the vertices belonging to the thresholded
set $V^+$. To summarize the process of merging, it suffices to take
into account that the new subgraphs are the connected components that
remain when nodes $v \notin V^+$ (and their edges) are excluded from
$G$.

\begin{figure}[!htb]
  \centering
  \includegraphics[width=0.9\columnwidth]{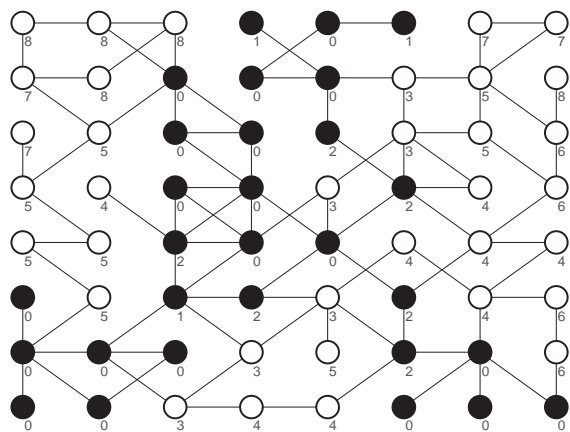}
  \caption{Merging $G^+$ inside $G$ (see Figure~\ref{fig:net}) for a
  threshold $T=2$. The value next to each node is its relevance
  $r(v)$, and the nodes of the subgraph merging are those shown in
  black.}  \label{fig:net_threshold}
\end{figure}

Thresholding with $T=0$ gives the original subgraphs, as $r(v)$ is
equal to zero if and only if $v$ belongs to some subgraph $G_i$. Since
greater thresholds include other nodes, two or more subgraphs can then
be joined into one single connected component. An example of a merging
for a threshold $T=2$ is given in Figure~\ref{fig:net_threshold},
where the input for the merging is the graph $G$ and its subgraphs
depicted in Figure~\ref{fig:net}. Figure~\ref{fig:net_threshold} shows
in black the nodes that belong to the merging, which results in a
single connected component joining all four subgraphs of $G$.

It is worth pointing out that recursive dilations could be used as a
method of subgraph merging (and not only as an intermediate step),
where each subgraph would be simulateneously dilated until the entire
network was covered. Nevertheless, Figure~\ref{fig:net_dilation} shows
that the first dilation of the subgraph $G_1$ includes nodes~14
and~24, whereas our merging does not include any of them until $T=3$
(see Figure~\ref{fig:net_threshold}). Although nodes~14 and~24 are
neighbors of $G_1$, they are farther from other subgraphs, and thus
they do not participate in short paths linking a pair of
subgraphs. This comparison shows that dilations do not discriminate
the relative position of nodes between subgraphs, and this is the
reason why we have defined the matrix of distances $D_{\delta}$ and
the relevance values $r(v)$.

If the merging is computed for sequentially increasing thresholds
starting at $T=0$, $V^+$ grows until $V^+ = V$, i.e. the subgraphs
$G_{i}$ expand until they form a single connected subgraph that is
equal to $G$, which we call \emph{gradual merging}. The number of
subgraphs (or connected components) $C^+$ in the merging can be
monitored until the end of the sequential thresholding, when $C^+$
must be equal to one. In this case, $C^+$ is a monotonically
nonincreasing function of $T$.

In a gradual merging, it is possible to verify the overall proximity
of the original subgraphs by observing how fast $C^+$ drops to one,
thus complementing the distance histograms explained in
Section~\ref{sec:hist}. Furthermore, the number of nodes in $V^+$, for
sequentially increasing thresholds, is useful to assess the overall
relevance $r(v)$ of nodes and also to measure how many nodes are
necessary to bring together all the original subgraphs. In order to
properly explain the utilization of gradual expansion and to
illustrate its potential to complement the distance histogram, we give
in the next section examples of subgraph characterization in both
real-world and artificial networks.


\section{Experimental Results and Discussion}
\label{sec:results}

We now illustrate the application of subgraph characterization to a
set of artificial and real-world networks. As already mentioned in
Section~\ref{sec:models}, 100 realizations of models ER, WS, BA and GG
were obtained, each one with $N=1,000$ nodes and mean degree
$\left\langle k \right\rangle = 6$. The real networks were introduced
in Section~\ref{sec:realnets}, namely NetScience, Email, Power Grid
and Internet-AS, with $N$ ranging from $379$ to $22,963$ and
$\left\langle k \right\rangle$ between $2.67$ and $9.62$ (please,
refer to Table~\ref{tab:realnets} for more details about real
networks). The chosen networks cover a considerable range of types of
structures usually studied in the field of complex networks
\cite{Albert2002,Dorogovtsev2002,Newman2003,Boccaletti2006,Costa2005a},
therefore providing a representative basis for the illustration of our
method.

An important step is the definition of the subgraphs to be
analyzed. To perform this task, we have chosen the measurements 
(i)~betweenness centrality $b$ and (ii)~clustering coefficient $c$ 
(both explained in Section~\ref{sec:measur}). More
specifically, the nodes with the highest $b$ or $c$ were chosen to
form subgraphs $G_i$ according to the definition presented in
Section~\ref{sec:charac}. Thus, subgraphs $G_i$ correspond to the 
connected components (sometimes containing only one node) existing between 
the nodes with highest betweenness centrality (or clustering
coefficient), limited to 2.5\% of the total number of nodes. The use
of betweenness centrality, already mentioned in the introductory
section of this paper, is particularly important in which concerns
proximity between groups of critical nodes. In other words, if these
subgraphs are close to each other, the network may become particularly
sensitive to directed attacks on central nodes. Subgraphs with high
clustering coefficient are also interesting to analyze because they
tend to be more cohesive than others, i.e. showing a variety of
different paths between its nodes. The characterization of the
connectivity between dense subgraphs may lead to the improvement of
search and transport strategies and also of network designs.  It is
important to observe that the analysis of the overall distribution of
the critical subgraphs through the network can provide complementary
information to the already important insights provided by those
measurements at the local topological level, as done traditionally.

In the next subsections we report the results obtained for the
aforementioned artificial and real-world networks regarding distance
histograms and subgraph merging. Since 100 realizations of each
network model have been performed, the following results are presented
in terms of average measurements and respective standard
deviations. Notice that for the real networks this procedure was not
necessary because only one network was available for each case.

\subsection{Network Models}
\label{sec:res_models}

\begin{figure*}[!htb]
	\centering
		\includegraphics[width=0.8\textwidth]{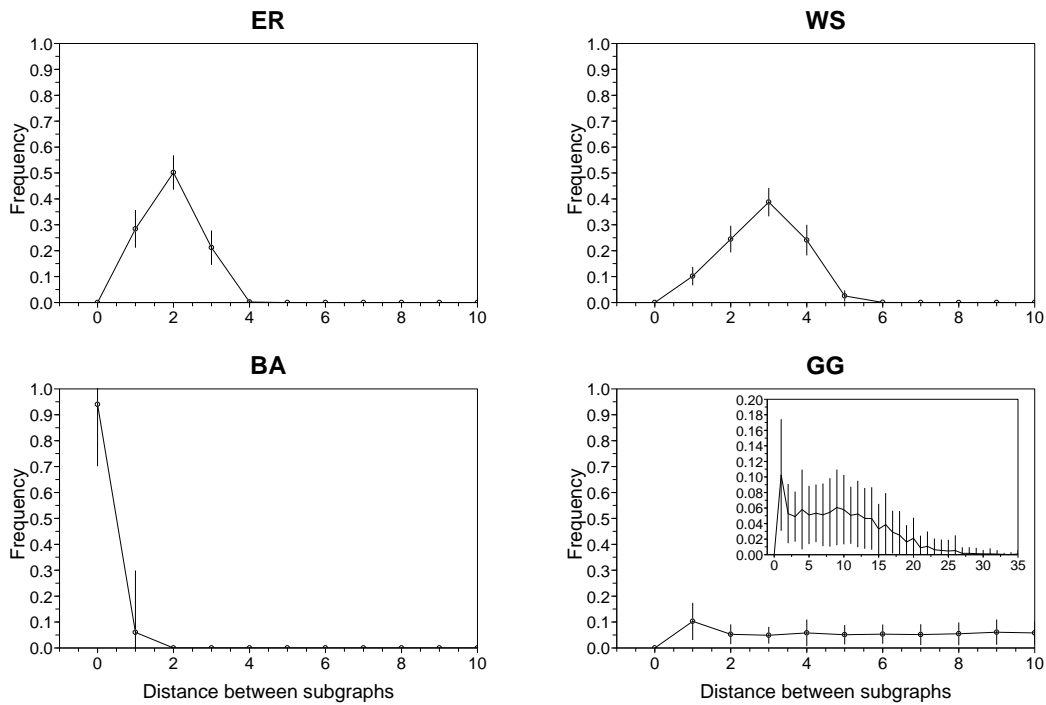}
	\caption{Average distance histograms and respective standard
	deviations considering 100 realizations of each network model
	(ER, WS, BA and GG). The subgraphs were created using the
	nodes (2.5\% of $N$) with the \emph{highest betweenness
	centrality}. }
	\label{fig:dists-modelnets-BC}
\end{figure*}

Figure~\ref{fig:dists-modelnets-BC} shows the average distance
histograms for the subgraphs consisting of nodes with the highest
betweenness centrality in ER, WS, BA and GG networks. Models ER and WS
have well-defined peaks at distance~2 and~3, respectively. Moreover,
distance values do not greatly deviate from the respective peaks in
both histograms, which shows that groups of central nodes tend to be
close one another in these two models. Interestingly, the BA model
shows a peak at distance~0, i.e. subgraphs are likely to be in the
same connected component in this model. Thus, ER, WS and BA models
have central subgraphs similar one another, although at different
intensities.  This ``central region'' plays an important role in a
network if we consider that the betweenness centrality measurement
reflects well the importance of nodes in dynamical processes taking
place in the network. For instance, diseases (or news) may spread fast
in a social network (typically well-modeled by WS) if the first
infected (or informed) people are inside the central
region. Procedures to stop epidemics may also have a higher success if
applied mostly at the central region. If we consider transport
processes, the central region needs to deal with considerable higher
traffic than the rest of the network and also needs to have stronger
security policies against attacks, otherwise a critical bottleneck may
arise. A very different behavior is shown by the model GG, with
distance values as high as~35. In general, subgraphs in geographical
networks are more likely to be distant 1-10 nodes apart, with lower
probabilities for higher distances. Although the histogram for the GG
model is not uniform, subgraphs consisting of nodes with high
betweenness centrality tend to be scattered over GG networks. Thus,
these networks do not have a main bottleneck since traffic would be
decentralized. Moreover, spreading processes would also be slower in
GG networks than in networks of ER, WS and BA types.

\begin{figure*}[!htb]
	\centering
		\includegraphics[width=0.8\textwidth]{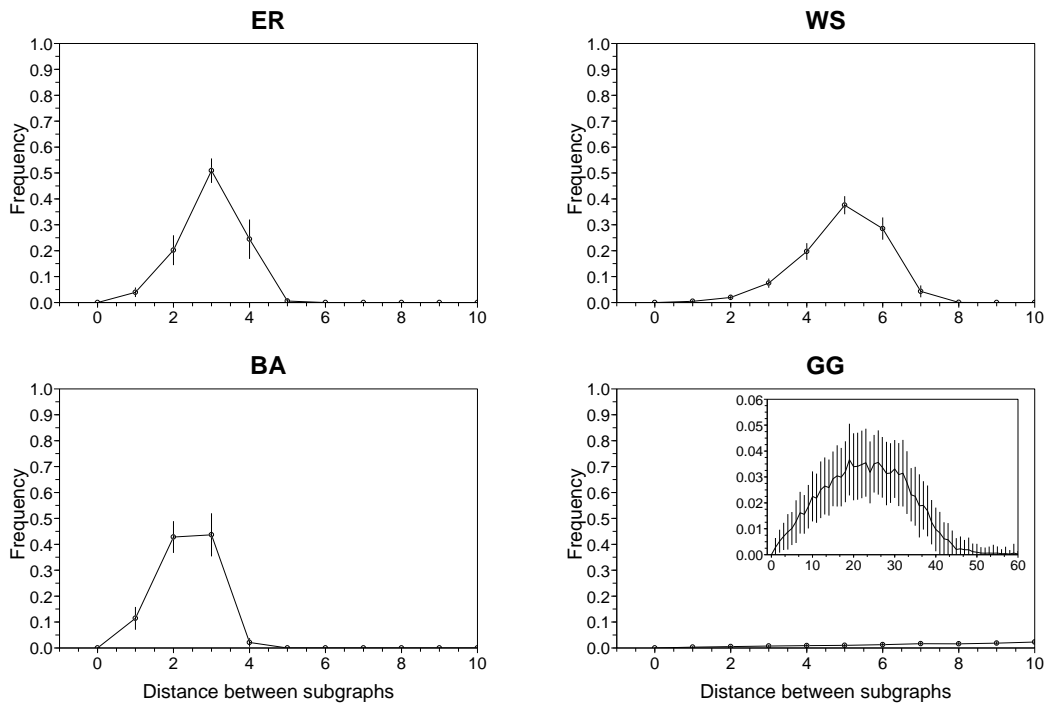}
	\caption{Average distance histograms and respective standard
	deviations considering 100 realizations of each network
	model. The subgraphs were created using the nodes (2.5\% of
	$N$) with the \emph{highest clustering coefficient}.}
	\label{fig:dists-modelnets-CC}
\end{figure*}

The histograms reproduced in Figure~\ref{fig:dists-modelnets-CC} were
obtained by taking nodes with high clustering coefficient as
references. These subgraphs are farther from each other than the
subgraphs created using betweenness centrality when considering a
comparison between the same network models. For instance, the models
ER and WS had their distance peaks increased from~2 to~3 and from~3
to~5, respectively, when comparing histograms of
Figures~\ref{fig:dists-modelnets-BC} with those of
Figure~\ref{fig:dists-modelnets-CC}. Nevertheless, these changes are
relatively small when considering the total number of nodes in these
networks ($N=1,000$). Subgraphs in BA networks now tend to be~2 or~3
units of distance apart, rather than being all connected as in the
previous BA histogram. Nevertheless, clustered subgraphs can be
considered close to each other in models ER, WS and BA, given the much
larger size of the networks when compared to the values in the
distance histograms. More significant increases in subgraph distances
were observed in the geographical model, where subgraphs were found to
be as far as~60 nodes apart. Now, a peak around 20-30 was found in the
distance histogram of the GG model, with a slow decay for higher
distance values. Clustered subgraphs can be regarded as a group of
nodes with redundancy of connections, since many paths exist between
two nodes in the same cluster. Thus, ER, WS and BA networks show link
redundancy at nearby locations, which can be an undesirable bottleneck
for transport networks in case of problems with this ``clustered
region''. Moreover, random failures happening inside low clustered
regions would isolate some nodes that do not have connection
redundancy around them. GG networks, on the other hand, have their
clustered subgraphs more dispersed, implying that connectivity
redundancy is not concentrated in a single region of the network. We
argue here that GG networks would then be more tolerant to localized
attacks since clustered subgraphs are spread all over the network.

\begin{figure*}[!htb]
	\centering
		\includegraphics[width=0.8\textwidth]{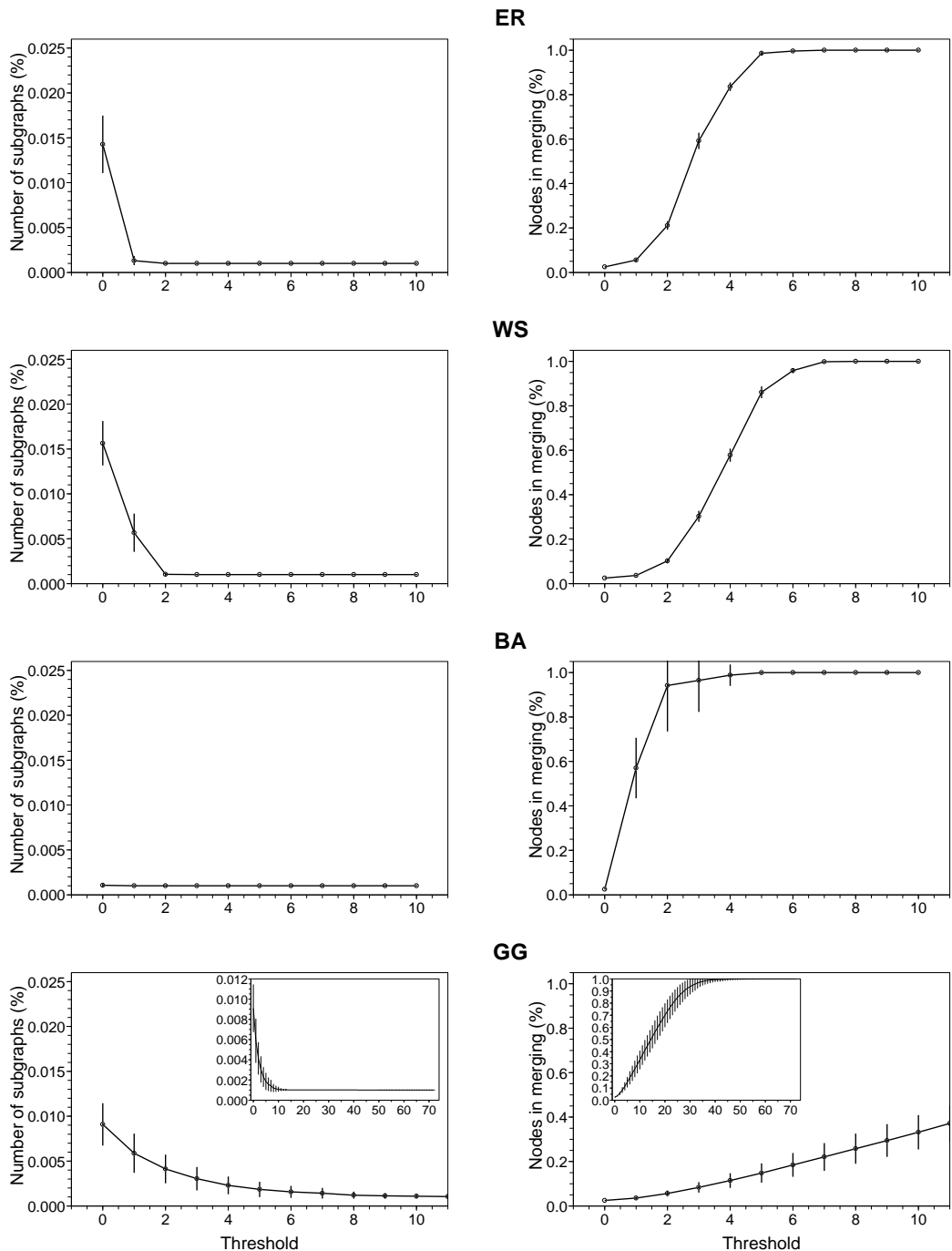}
	\caption{Average number of subgraphs $C^+$ (left panel) and
	average number of nodes $|V^+|$ (right panel) in the gradual
	merging performed for 100 realizations of models ER, WS, BA
	and GG (standard deviations are denoted by vertical
	bars). Values in the vertical axes are relative to the total
	number of nodes in these networks. The subgraphs were created
	using the nodes (2.5\% of $N$) with the \emph{highest
	betweenness centrality}.}  \label{fig:res-modelnets-BC}
\end{figure*}

\begin{figure*}[!htb]
	\centering
		\includegraphics[width=0.8\textwidth]{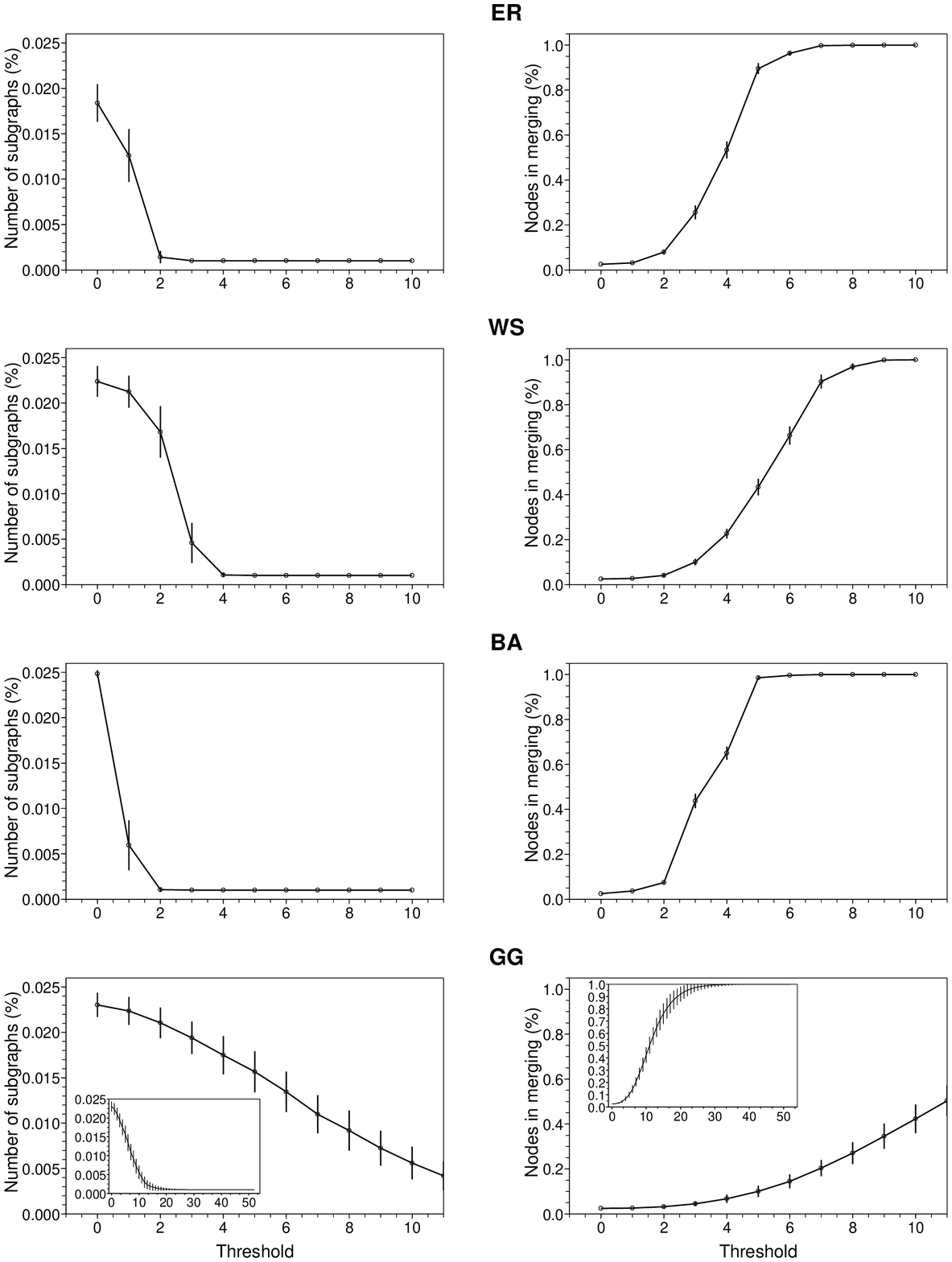}
	\caption{Average number of subgraphs $C^+$ (left panel) and
	average number of nodes $|V^+|$ (right panel) in the gradual
	merging performed for 100 realizations of models ER, WS, BA
	and GG. The subgraphs were created using the nodes (2.5\% of
	$N$) with the \emph{highest clustering coefficient}.}
	\label{fig:res-modelnets-CC}
\end{figure*}

We now turn our attention to the gradual merging of subgraphs. Our
approach consists in monitoring the number of subgraphs $C^+$ and the
number of nodes $|V^+|$ in the gradual merging as threshold $T$
increases. The plots in the left column of
Figures~\ref{fig:res-modelnets-BC} and~\ref{fig:res-modelnets-CC}
show, for each network model, the number of subgraphs $C^+$ as a
function of the merging threshold $T$, while the right column shows
the number of nodes $|V^+|$ as a function of $T$. Both quantities were
divided by the total number of nodes $N$ in the network, thus
normalizing their range throughout this paper. Observe also that when
the $C^+$ curve stabilizes, its absolute value becomes equal to one.

Figure~\ref{fig:res-modelnets-BC} shows the results for network models
with subgraphs derived from the betweenness centrality
measurement. The plots obtained for the ER model (first line of
Figure~\ref{fig:res-modelnets-BC}) indicate that a threshold $T=1$ is
capable of joining almost all subgraphs using approximately $|V^+| =
5\%$ of the nodes in the network, including the nodes inside
subgraphs. These results show that subgraphs with central nodes tend
to be close one another in ER networks, a feature already noticed in
the analysis of distance histograms. Nevertheless, with the gradual
merging we are able to identify which (and how many) nodes are more
relevant while joining all subgraphs. Thus, only 5\% of the nodes in
ER networks is enough to group its central subgraphs in one connected
component, reinforcing the idea of a ``central region'' introduced in
the beginning of this subsection. WS networks show similar results,
where all subgraphs are merged when $T=2$ using approximately $|V^+| =
10\%$ of network nodes. The central region is more prominent in BA
networks because nodes with high betweenness centrality tend to form a
single connected component from the onset of the gradual merging
(which, by the way, behaves simply as a dilation in this
case). Geographical networks, on the other hand, need a threshold
$T=13$ to connect all subgraphs, encompassing approximately $|V^+| =
45\%$ of network nodes. Indeed, as already observed in this
subsection, subgraphs consisting of central nodes are likely to be
distant from each other in GG networks, thus the high threshold and
$|V^+|$ necessary to connect all subgraphs.

Figure~\ref{fig:res-modelnets-CC} shows the number of subgraphs $C^+$
and the number of nodes $|V^+|$ in the gradual merging of subgraphs
with high clustering coefficient. The left panel of this figure shows
that the chosen subgraphs tend to be composed of single nodes,
specially in BA networks (notice that the maximum number of subgraphs
in this case is 2.5\% of $N$, when all subgraphs are
unitary). Moreover, for each network model the threshold necessary to
join all subgraphs is consistently higher than that observed in
Figure~\ref{fig:res-modelnets-BC}, showing again that clustered
subgraphs are farther from each other than central subgraphs. More
detailed features of Figure~\ref{fig:res-modelnets-CC} were also
noticed: ER and BA networks have $C^+ = 1$ when $T=2$ ($C^+$ drops
faster in BA than in ER), using less than 10\% of network
nodes. Subgraphs in WS networks are all merged when $T=4$, with more
than 20\% of network nodes included in the merging. Geographical
networks show again the most distinct results: a threshold $T \approx
25$ is necessary to bring together all subgraphs, when almost the
entire network (more than 90\% of the nodes) is included in the
merging.

The results reported in this subsection show a remarkable difference
between the GG model and the other three models. Geographical networks
demand considerable higher thresholds to merge all subgraphs, possibly
a consequence of the distance parameter that controls the creation of
edges. The BA model shows a distinctive feature for highly central
subgraphs: they are, in fact, a single connected component that groups
all nodes with high betweenness centrality. These nodes are likely to
be the hubs, since betweenness centrality and degree were shown to be
highly correlated in BA networks \cite{Holme2002}. On the other hand,
highly clustered nodes are apart from each other in the BA model,
which indicates that these nodes tend to approximate the periphery of
BA networks. ER and WS networks show similar results, although in the
WS model subgraphs are consistently farther away from each other than
in the ER model. Notice that both models have low average shortest
paths; nevertheless, the WS model has higher average clustering
coefficient than the ER model, which may indicate the reason for the
observed differences in their results.

\subsection{Real-World Networks}
\label{sec:res_realnets}

\begin{figure*}[!htb]
	\centering
		\includegraphics[width=0.7\textwidth]{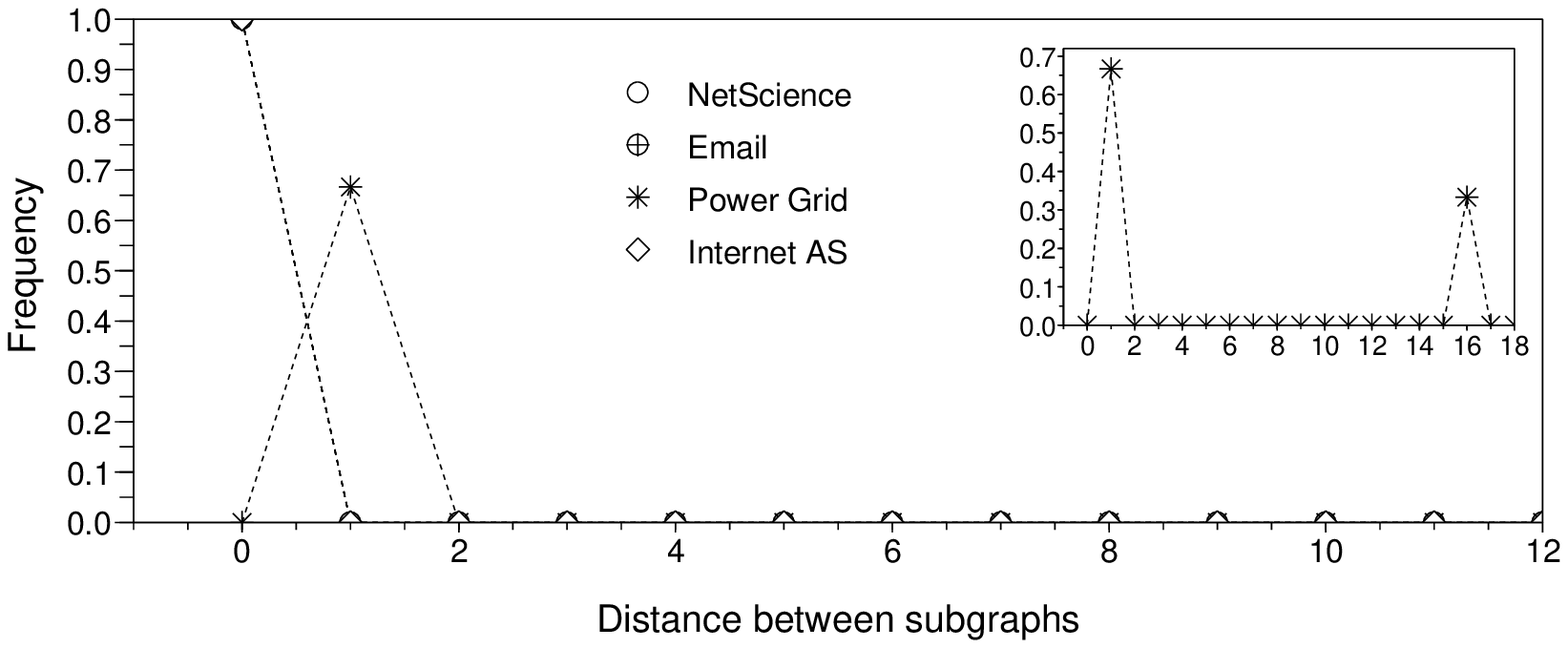}
	\caption{Distance histograms for real networks. The subgraphs
	were created using the nodes (2.5\% of $N$) with the
	\emph{highest betweenness centrality}.}
	\label{fig:dists-realnets-BC}
\end{figure*}

The distance histograms for real networks considering subgraphs
derived from the betweenness centrality measurement are shown in
Figure~\ref{fig:dists-realnets-BC}. Three networks (NetScience, Email
and Internet-AS) have central nodes contained in the same subgraph,
thus only distance~0 is counted in their histograms. This observation
may indicate a weakness in Internet's architecture: although Internet
is not centrally controlled, some important autonomous systems
(according to the betweenness centrality) are physically connected to
each other, thus forming a central group of nodes that may corrupt the
entire network if attacked. The NetScience and Email networks show the
same distance distribution, which means that a small group of people
plays an important role concentrating a considerable amount of
knowledge/information flow among nodes. Subgraphs in the Power Grid
network are slightly more separated from each other. Nevertheless,
almost 70\% of its subgraph distances are equal to~1, indicating that
the majority of central subgraphs are close to each other in this
network, which may be considered a security flaw in the Power Grid.

\begin{figure*}[!htb]
	\centering
		\includegraphics[width=0.7\textwidth]{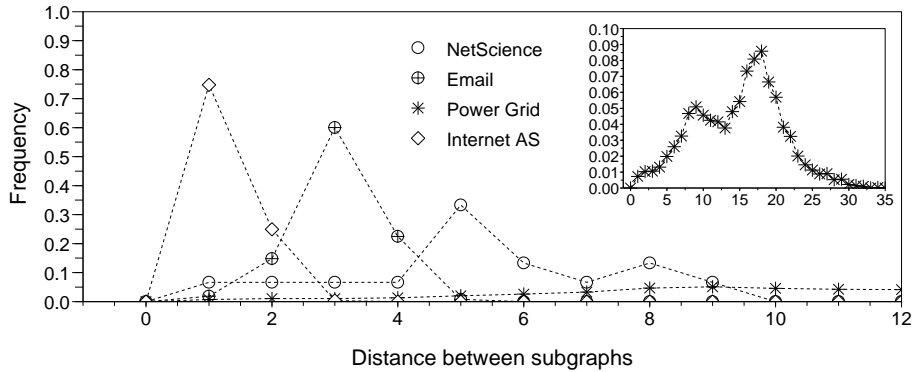}
	\caption{Distance histograms for real networks. The subgraphs
	were created using the nodes (2.5\% of $N$) with the
	\emph{highest clustering coefficient}.}
	\label{fig:dists-realnets-CC}
\end{figure*}

As for models ER, WS, BA and GG (Section~\ref{sec:res_models}), 
real networks have clustered nodes more scattered over the 
network than central nodes (Figure~\ref{fig:dists-realnets-CC}
shows the distance histograms for subgraphs based on high clustering
coefficient). At one extremity is the Power Grid, with subgraphs
distant at most 33 units of distance, with peaks at distances~9
and~18. This fact suggests a good tolerance to random failures, since
the network has redundancy of connections spread over the 
network. At the opposite side is the Internet, with all clustered
subgraphs near one another (at most with distance~2). We argue here 
that autonomous systems should have more distributed clustered
subgraphs in order to avoid bottlenecks at regions far away from the
observed clustered regions. The NetScience and Email networks have 
distance peaks at~5 and~3, respectively, indicating that clusters
of collaborators/acquaintances are not too distant from each other 
in these networks.

\begin{figure*}[!htb]
	\centering
		\includegraphics[width=0.7\textwidth]{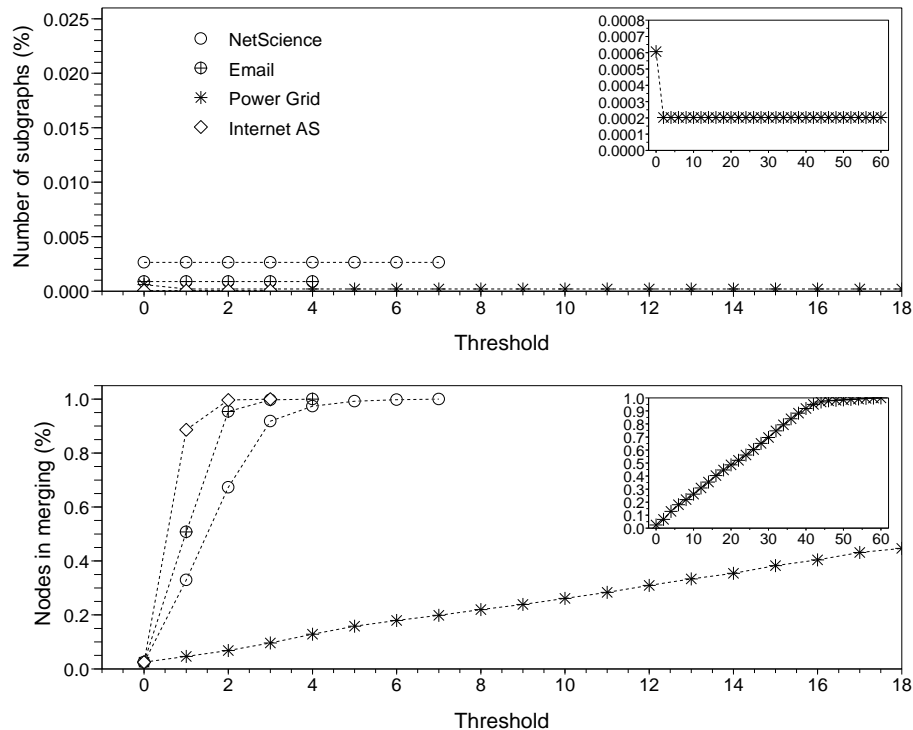}
	\caption{Number of subgraphs $C^+$ (top plot) and number of
	nodes $|V^+|$ (bottom plot) in the gradual merging performed
	for the real networks. Values in the vertical axes are
	relative to the total number of nodes in these networks. The
	subgraphs were created with respect to the nodes (2.5\% of $N$) 
        with the
	\emph{highest betweenness centrality}.}
	\label{fig:res-realnets-BC}
\end{figure*}

Figure~\ref{fig:res-realnets-BC} depicts the gradual merging resulting
in real networks with subgraphs constructed using the betweenness
centrality measurement. As already mentioned in this section,
NetScience, Email and Internet networks have one single central
subgraph, i.e. $C^+ = 1$ for any merging threshold. Thus, their
merging acts as a dilation with a steep increase of $|V^+|$. The Power
Grid has only a few subgraphs joined at $T=1$, using 5\% of
network nodes. Nevertheless, the entire Power Grid network is only
covered by the merging when $T \approx 55$, which is in accordance
with the slow growth of $|V^+|$ observed for geographical networks in
Section~\ref{sec:res_models}.

\begin{figure*}[!htb]
	\centering
		\includegraphics[width=0.7\textwidth]{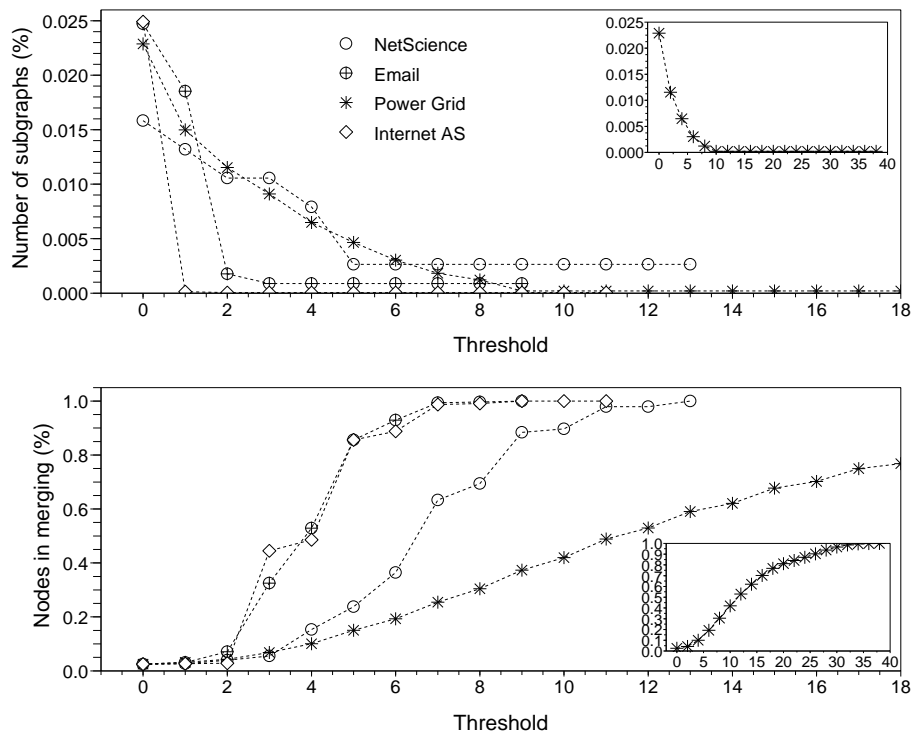}
	\caption{Number of subgraphs $C^+$ (top plot) and number of
	nodes $|V^+|$ (bottom plot) in the gradual merging performed
	for the real networks. The subgraphs were created using the
	nodes (2.5\% of $N$) with the \emph{highest clustering
	coefficient}.}  \label{fig:res-realnets-CC}
\end{figure*}

Rather different results were found for subgraphs based on nodes with
high clustering coefficient (see
Figure~\ref{fig:res-realnets-CC}). High clustered nodes tend to be
separated from each other in these real networks, as already observed
for models ER, WS, BA and GG (i.e. $C^+ \rightarrow 2.5\%$ when $T=0$,
specially for Email, Power Grid and Internet networks, showing that
almost every subgraph is composed of single nodes).  Remarkably, less
than 5\% of Internet nodes are capable of joining all its subgraphs at
$T=2$. Subgraphs in the Email network are also quickly joined (at
$T=3$), although demanding almost 35\% of its nodes. NetScience
maintains its subgraphs separated until $T=5$, when almost 25\% of its
nodes are merged. These observations show that, although Internet,
Email and NetScience have subgraphs quickly merged, they are more
cohesive in the Internet because much less nodes are necessary to
bring them together in a single connected component. The Power Grid
only joins its subgraphs when $T=9$, which is again in accordance with
previous results (Section~\ref{sec:res_models}) that show a slow
merging in GG networks.

Although the Power Grid can be regarded as a geographical network, it
was previously associated with the WS model because it shows the
small-world effect \cite{Watts1998}. Nevertheless, in the experiments
reported here the Power Grid shows rather different behaviors than the
observed in the WS model. The Internet, although being geographically
constrained, shows very different results than model GG in all
experiments we have performed. In fact, the Internet is better
associated with the BA model, since it shows a power-law degree
distribution~\cite{Faloutsos1999}. Indeed, the Internet at the
autonomous system level shows results similar to the ones obtained for
the BA model: central nodes are all connected in only one subgraph and
clustered nodes are more scattered over the network.  NetScience and Email
networks also present a behavior similar to the observed in the BA
model. NetScience can indeed be associated with the BA model, as
power-law degree distributions were observed in scientific
collaboration networks~\cite{Newman2001a}, although the Email network
deviates from this model by having an exponential distribution of
degrees~\cite{Guimera2003}.


\section{Concluding Remarks}
\label{sec:concl}

In this paper we presented a framework for characterizing the
distribution of critical subgraphs in complex networks. We adopted
distance histograms to assess the overall relationship between
subgraphs and also developed an algorithm to sequentially merge
subgraphs according to a metric of node relevance. The merging
approach complements the distance histogram by identifying which (and
how many) nodes are necessary to join two or more subgraphs in the
same connected component.

Rather than characterizing single nodes exclusively, the proposed
framework operates at a higher topological level by analyzing groups
of nodes and their interconnectivity. Closely related topological
levels have been the focus of many network-based studies, such as the
analysis of communities and motifs. Nevertheless, differently than
communities, the method proposed in this paper does not create a
partition of the network, and and also does not identify small
subgraph patterns (i.e. motifs). Our main motivation is to analyze the
interconnectivity and dispersion of similar (according to any desired
criteria) groups of nodes independently of their size.

We illustrated our method by analyzing critical subgraphs with respect
to both theoretical and real-world networks. Subgraphs comprising
nodes with high betweenness centrality were found to be close one
another in models Erd\H{o}s-R\'{e}nyi, Watts-Strogatz and
Barab\'{a}si-Albert, as well as in the following real-world networks:
Email, NetScience and Internet-AS.  All these networks also presented
clustered subgraphs (i.e. with nodes with high clustering coefficient)
close to each other, although a bit farther than the subgraphs based
on the centrality measurement. Furthermore, both types of subgraphs
were found to be more distant one another in the Geographical model
and also in the Power Grid network. The experimental findings reported
in this paper contribute to a better understanding of the structure of
the aforementioned networks, allowing us to draw some conclusions
about dynamical processes taking place on networks.

Further work may focus on similar analysis using other networks, as
well as different types of subgraphs. Another interesting
investigation would be to expand the proposed framework using
hierarchical/concentric measurements
\cite{Costa2004b}, thus allowing the analysis of subgraph neighborhood at 
different hierarchical levels.


\begin{acknowledgments}
L.~da~F. Costa is grateful to FAPESP (05/00587-5) and CNPq
(301303/06-1) for financial support. L.~Antiqueira thanks the
sponsorship of FAPESP (06/61743-7). The authors also would like to
thank M.E.J.~Newman, A.~Arenas and D.~Watts for making their datasets
available on the WWW.
\end{acknowledgments}

\bibliographystyle{apsrev}
\bibliography{paper_subgraphs}

\end{document}